\documentclass[a4paper,12pt]{article}
\usepackage{amsfonts}
\usepackage{amsmath}

\input{tcilatex}

\begin{document}

\title{ Pseudo--Hermiticity and weak pseudo--Hermiticity: Equivalence of
complementarity and the coordinate transformations in position--dependent
mass}
\author{S.--A.\ Yahiaoui\thanks{%
E-mail: s\_yahiaoui@mail.univ-blida.dz}, M.\ Bentaiba\thanks{%
E-mail:\ bentaiba@mail.univ-blida.dz} \\
%EndAName
LPTHIRM, D\'{e}partement de physique, Facult\'{e} des Sciences,\\
Universit\'{e} Saad DAHLAB de Blida, Blida, Algeria}
\maketitle

\begin{abstract}
The complementarity between the twin concepts of pseudo--Hermiticity and
weak pseudo--Hermiticity, established by Bagchi and Quesne [Phys. Lett. A 
\textbf{301 }(2002) 173-176], can be understood in terms of coordinate
transformations.


\end{abstract}

\section{Introduction}

In recent years, the concept of pseudo--Hermiticity has attracted much
attention on behalf of physicists [1-9]. The basic mathematical structure
underlying the properties of pseudo-Hermiticity is revealed [3-5] and it has
been found to be a more general concept then those of Hermiticity and $%
\mathcal{PT}$--symmetry [10-15]. By definition, a linear operator $H$ (here
a Hamiltonian) acting in a Hilbert space $\mathfrak{H}$ is called $\eta $%
--pseudo--Hermitian if it obeys to [3-5]%
\begin{equation}
\eta H=H^{\dag }\eta ,  \tag{1}
\end{equation}%
where $\eta $ is a Hermitian linear invertible operator and a dagger stands
for the adjoint of the corresponding operator. Then (non--Hermitian)
Hamiltonian $H$ has a real spectrum [3] if there is an invertible linear
operator $d:$ $\mathfrak{H\rightarrow H}$ such that $\eta =d^{\dag }d$. As a
consequence of this, the reality of the bound--state eigenvalues of $H$ can
be associated with $\eta $--pseudo--Hermiticity. Note that choosing $\eta =1$
reduces the assumption (1) to the Hermiticity.

In a very interesting work [7], Bagchi and Quesne point out that the twin
concepts of pseudo--Hermiticity and weak pseudo--Hermiticity are
complementary to one another by admitting that it is possible to break up $%
\eta $ into two operators, i.e. $\eta _{+}$ and $\eta _{-}$, following
combinations%
\begin{equation}
\eta _{+}H=H^{\dag }\eta _{+}\qquad \text{and\qquad }\eta _{-}H=H^{\dag
}\eta _{-},  \tag{2}
\end{equation}%
where $\eta _{\pm }=\eta \pm \eta ^{\dag }$. The first assumption
corresponds to the pseudo--Hermiticity where $\eta _{+}$ is a second--order
differential realization while the second is associated with weak
pseudo--Hermiticity and $\eta _{-}$ is a first--order realization.

In the present paper, we take up the study of a complementarity between
pseudo--Hermiticity and weak pseudo--Hermiticity under the concept of
coordinate transformation and examine how the pseudo--Hermiticity should map
to the weak pseudo--Hermiticity. In fact, our primary concern is to point
out that the coordinate transformations can be looked upon as a toy model
for understanding the complementarity. In this light, the complementarity
acquires a mathematical meaning which, unfortunately, was not established in
[7].

We end this section by defining a quite formalism used throughout the
present work. In the case of a spatially varying mass [16-19] which will be
denoted by $M\left( x\right) =m_{0}m\left( x\right) $, the Hamiltonian
proposed by von Roos [16] reads%
\begin{equation}
H=\frac{1}{4}\left( m^{\alpha }\left( x\right) pm^{\beta }\left( x\right)
pm^{\gamma }\left( x\right) +m^{\gamma }\left( x\right) pm^{\beta }\left(
x\right) pm^{\alpha }\left( x\right) \right) +V\left( x\right) ,  \tag{3}
\end{equation}%
where $\alpha $, $\beta $ and $\gamma $ are three parameters which obey to
the relation $\alpha +\beta +\gamma =-1$ in order to grant the classical
limit and $V\left( x\right) =V_{\func{Re}}\left( x\right) +iV_{\func{Im}%
}\left( x\right) \in 
%TCIMACRO{\U{2102} }%
%BeginExpansion
\mathbb{C}
%EndExpansion
$. Here, $p\left( =-i\frac{d}{dx}\right) $ is a momentum with $\hbar
=m_{0}=1 $, and $m\left( x\right) $ is dimensionless--real valued mass.

Using the restricted Hamiltonian from the $\alpha =\gamma =0$ and $\beta =-1$
constraints [17], the Hamiltonian (3) becomes%
\begin{equation}
H=pU^{2}\left( x\right) p+V\left( x\right) ,  \tag{4}
\end{equation}%
with $U^{2}\left( x\right) =\frac{1}{2m\left( x\right) }$ and $U\left(
x\right) \in 
%TCIMACRO{\U{211d} }%
%BeginExpansion
\mathbb{R}
%EndExpansion
$. The shift on the momentum $p$ in the manner%
\begin{equation}
p\rightarrow p^{\prime }=p-\frac{A\left( x\right) }{U\left( x\right) }, 
\tag{5}
\end{equation}%
where $A\left( x\right) =a\left( x\right) +ib\left( x\right) \in 
%TCIMACRO{\U{2102} }%
%BeginExpansion
\mathbb{C}
%EndExpansion
$ and $a\left( x\right) $, $b\left( x\right) $ are real functions, allows to
bring the Hamiltonian (4) in the form%
\begin{equation}
H\rightarrow \mathcal{H}=\left( p-\frac{A\left( x\right) }{U\left( x\right) }%
\right) U^{2}\left( x\right) \left( p-\frac{A\left( x\right) }{U\left(
x\right) }\right) +V\left( x\right) .  \tag{6}
\end{equation}%
\noindent

\section{Pseudo--Hermiticity generating function}

As $\eta _{+}\left( =d^{\dag }d\right) $ is pseudo--Hermitian and following
the ordinary supersymmetric quantum mechanics, the operators $d$ and $%
d^{\dag }$ are connecting to the first--order differential realization
through [8,9]%
\begin{eqnarray}
d &=&U\left( x\right) \frac{d}{dx}+\Phi \left( x\right) ,  \TCItag{7.a} \\
d^{\dag } &=&-U^{\prime }\left( x\right) -U\left( x\right) \frac{d}{dx}+\Phi
^{\ast }\left( x\right) ,  \TCItag{7.b}
\end{eqnarray}%
where $\Phi \left( x\right) =F\left( x\right) +iG\left( x\right) \in 
%TCIMACRO{\U{2102} }%
%BeginExpansion
\mathbb{C}
%EndExpansion
$ and $F\left( x\right) $, $G\left( x\right) $ are real functions. Here, the
prime denotes derivative with respect to $x$. It is obvious that
Eqs.(7.a--b) become, under the transformation (5),%
\begin{eqnarray}
d &\rightarrow &\mathcal{D}=U\left( x\right) \frac{d}{dx}-iA\left( x\right)
+\Phi \left( x\right) ,  \TCItag{8.a} \\
d^{\dag } &\rightarrow &\mathcal{D}^{\dag }=-U^{\prime }\left( x\right)
-U\left( x\right) \frac{d}{dx}+iA^{\ast }\left( x\right) +\Phi ^{\ast
}\left( x\right) ,  \TCItag{8.b}
\end{eqnarray}%
and in terms of these, $\eta _{+}$ is transformed into $\widetilde{\eta }%
_{+}\left( =\mathcal{D}^{\dag }\mathcal{D}\right) $ such as%
\begin{equation}
\widetilde{\eta }_{+}=-U^{2}\left( x\right) \frac{d^{2}}{dx^{2}}-2\mathcal{K}%
\left( x\right) \frac{d}{dx}+\mathcal{L}\left( x\right) ,  \tag{9}
\end{equation}%
where $\mathcal{K}\left( x\right) $ and $\mathcal{L}\left( x\right) $ are
defined as%
\begin{eqnarray}
\mathcal{K}\left( x\right) &=&U\left( x\right) U^{\prime }\left( x\right)
+iU\left( x\right) \left( G\left( x\right) -a\left( x\right) \right) , 
\TCItag{10.a} \\
\mathcal{L}\left( x\right) &=&\Phi ^{\ast }\left( x\right) \Phi \left(
x\right) +A^{\ast }\left( x\right) A\left( x\right) -\left[ U\left( x\right)
\left( iA\left( x\right) -\Phi \left( x\right) \right) \right] ^{\prime } 
\notag \\
&&-i\Phi ^{\ast }\left( x\right) A\left( x\right) +i\Phi \left( x\right)
A^{\ast }\left( x\right) .  \TCItag{10.b}
\end{eqnarray}

Taking the adjoint of Eq.(9), one can easily check that $\widetilde{\eta }%
_{+}$ is Hermitian; since it is written in the form $\widetilde{\eta }_{+}=%
\mathcal{D}^{\dag }\mathcal{D}$. On the other hand, the Hamiltonian (6) may
be expressed as%
\begin{equation}
\mathcal{H}=-U^{2}\left( x\right) \frac{d^{2}}{dx^{2}}-2\mathcal{M}%
_{1}\left( x\right) \frac{d}{dx}+\mathcal{N}_{1}\left( x\right) +V\left(
x\right) ,  \tag{11}
\end{equation}%
where%
\begin{eqnarray}
\mathcal{M}_{1}\left( x\right) &=&U\left( x\right) U^{\prime }\left(
x\right) -iU\left( x\right) A\left( x\right) ,  \TCItag{12.a} \\
\mathcal{N}_{1}\left( x\right) &=&i\left[ U\left( x\right) A\left( x\right) %
\right] ^{\prime }+A^{2}\left( x\right) .  \TCItag{12.b}
\end{eqnarray}

It should be noted that $\mathcal{D}$ and $\mathcal{D}^{\dag }$ are two
intertwining operators, and then the defining assumption (1) can be
generalized into $\widetilde{\eta }_{+}\mathcal{H=H}^{\dag }\widetilde{\eta }%
_{+}$. Using Eqs.(9), (11) and the adjoint of Eq.(11) on both sides of the
last equation and comparing between their varying differential coefficients,
we can recognized from the third--derivative that $b\left( x\right) =0$,
while the second--derivative connects the potential to its conjugate through%
\begin{equation}
V\left( x\right) =V^{\ast }\left( x\right) -4iU\left( x\right) G^{\prime
}\left( x\right) .  \tag{13}
\end{equation}

However, the coefficients corresponding to the first--derivative give the
shape of the potential, where after integration, we get%
\begin{equation}
V\left( x\right) =F^{2}\left( x\right) -G^{2}\left( x\right) -\left[ U\left(
x\right) F\left( x\right) \right] ^{\prime }-2iU\left( x\right) G^{\prime
}\left( x\right) +\delta ,  \tag{14}
\end{equation}%
where $\delta $ is some constant of integration. The last remaining
coefficient corresponds to the null--derivative and gives the
pure--imaginary differential equation%
\begin{eqnarray}
F^{2}\left( x\right) -\left[ U\left( x\right) F\left( x\right) \right]
^{\prime } &=&\frac{G\left( x\right) }{G^{\prime }\left( x\right) }\left(
-F\left( x\right) F^{\prime }\left( x\right) +\frac{1}{2}\left[ U\left(
x\right) F\left( x\right) \right] ^{\prime \prime }\right)  \notag \\
&&+\frac{1}{G^{\prime }\left( x\right) }\left( \frac{1}{4}\left[ U^{2}\left(
x\right) G^{\prime \prime }\left( x\right) \right] ^{\prime }-\frac{G\left(
x\right) }{4}\left[ U\left( x\right) U^{\prime \prime }\left( x\right) %
\right] ^{\prime }\right.  \notag \\
&&+\left. \frac{U^{\prime }\left( x\right) U\left( x\right) }{4}\left[ \frac{%
G\left( x\right) }{U\left( x\right) }\right] ^{\prime \prime }+\frac{%
U^{\prime 2}\left( x\right) U\left( x\right) }{2}\left[ \frac{G\left(
x\right) }{U\left( x\right) }\right] ^{\prime }\right)  \notag \\
&&-\frac{U^{\prime \prime }\left( x\right) U\left( x\right) }{4}, 
\TCItag{15}
\end{eqnarray}%
which is not easy to solve. However, the $\widetilde{\eta }_{+}$%
--orthogonality suggests that the eigenvector, here $\Psi \left( x\right) $,
is related to $\mathcal{H}$ through%
\begin{equation}
\widetilde{\eta }_{+}\Psi \left( x\right) =0,\quad \text{or\quad }\mathcal{D}%
\Psi \left( x\right) =0,  \tag{16}
\end{equation}%
leading, after integration, to the ground--state wave function%
\begin{eqnarray}
\Psi \left( x\right) &=&\Lambda \left( x\right) \psi \left( x\right)  \notag
\\
&=&\exp \left[ i\int^{x}dy\frac{A\left( x\right) }{U\left( x\right) }\right]
\psi \left( x\right)  \notag \\
&=&\mathcal{N}_{0}\exp \left[ -i\int^{x}dy\frac{F\left( x\right) }{U\left(
x\right) }-i\int^{x}dy\frac{G\left( x\right) -a\left( x\right) }{U\left(
x\right) }\right] ,  \TCItag{17}
\end{eqnarray}%
where $\mathcal{N}_{0}$ is a constant of normalization. The wave function $%
\Psi \left( x\right) $ is then subjected to a gauge transformation in a
manner of $\psi \left( x\right) \rightarrow \Psi \left( x\right) =\Lambda
\left( x\right) \psi \left( x\right) $, where $\Lambda \left( x\right) =%
\sqrt{\widetilde{\eta }_{+}\left( x\right) }$ [1,7]. Now, using the Schr\"{o}%
dinger equation $\mathcal{H}\Psi \left( x\right) =\mathcal{E}\Psi \left(
x\right) $ where $\mathcal{E}=\mathcal{E}_{\func{Re}}+i\mathcal{E}_{\func{Im}%
}$, one obtain the differential equation%
\begin{equation}
2F\left( x\right) G\left( x\right) +U\left( x\right) G^{\prime }\left(
x\right) -U^{\prime }\left( x\right) G\left( x\right) =-\mathcal{E}_{\func{Im%
}}+i\left( \mathcal{E}_{\func{Re}}-\delta \right) ,  \tag{18}
\end{equation}%
where $\delta $ is a constant introduced in Eq.(14). In order to solve
suitably Eq.(18), we assume that both sides of Eq.(18) are equal to zero;
which requires that $\mathcal{E}_{\func{Re}}=\delta $ and $\mathcal{E}_{%
\func{Im}}=0$. Therefore, the energy eigenvalues $\mathcal{E}$ are real. In
these settings, we end up by relating $F\left( x\right) $ to $G\left(
x\right) $ and $U\left( x\right) $ through the differential equation%
\begin{equation}
F\left( x\right) =\frac{G\left( x\right) }{2}\left[ \frac{U\left( x\right) }{%
G\left( x\right) }\right] ^{\prime },  \tag{19}
\end{equation}%
and which proves to be the solution of Eq.(15). Hence, it becomes clear that 
$F\left( x\right) $ (i.e. $G\left( x\right) $) is a generating function
leading to identify the potential $V\left( x\right) $.

\section{Weak pseudo--Hermiticity generating function}

For the first--order differential realization, $\eta _{-}$ may be
anti--Hermitian and $\mathcal{H}$ can be relaxed to be weak
pseudo--Hermitian. Then $\eta _{-}$ can be expressed as%
\begin{equation}
\eta _{-}=U\left( x\right) \frac{d}{dx}+\varphi \left( x\right) ,  \tag{20}
\end{equation}%
where $\varphi \left( x\right) =f\left( x\right) +ig\left( x\right) \in 
%TCIMACRO{\U{2102} }%
%BeginExpansion
\mathbb{C}
%EndExpansion
$ and $f\left( x\right) $, $g\left( x\right) $ are real functions. Using
Eq.(5), $\eta _{-}$ and $\eta _{-}^{\dag }$ become%
\begin{eqnarray}
\eta _{-} &\rightarrow &\widetilde{\eta }_{-}=U\left( x\right) \frac{d}{dx}%
-iA\left( x\right) +\varphi \left( x\right) ,  \TCItag{21.a} \\
\eta _{-}^{\dag } &\rightarrow &\widetilde{\eta }_{-}^{\dag }=-U^{\prime
}\left( x\right) -U\left( x\right) \frac{d}{dx}+iA^{\ast }\left( x\right)
+\varphi ^{\ast }\left( x\right) .  \TCItag{21.b}
\end{eqnarray}

As now $\widetilde{\eta }_{-}$ points to weak pseudo--Hermiticity, this
amounts to writing%
\begin{equation}
\widetilde{\eta }_{-}^{\dag }=-\widetilde{\eta }_{-},  \tag{22}
\end{equation}%
which brings to the relation%
\begin{equation}
U^{\prime }\left( x\right) =2f\left( x\right) +2b\left( x\right) .  \tag{23}
\end{equation}

Letting both sides of $\widetilde{\eta }_{-}\mathcal{H=H}^{\dag }\widetilde{%
\eta }_{-}$ act on every function and comparing their varying differential
coefficients, one deduced from the second--derivative that $b\left( x\right)
=0$, therefore the generating function $f\left( x\right) $ in Eq.(23) becomes%
\begin{equation}
f\left( x\right) =\frac{U^{\prime }\left( x\right) }{2},  \tag{24}
\end{equation}%
while the first--derivative gives the imaginary part of the potential%
\begin{equation}
V_{\func{Im}}\left( x\right) =iU\left( x\right) f^{\prime }\left( x\right)
-U\left( x\right) g^{\prime }\left( x\right) -\frac{i}{2}U\left( x\right)
U^{\prime \prime }\left( x\right) .  \tag{25}
\end{equation}

The last coefficient corresponds to the null--derivative which gives, after
a double integration by parts, the real part of the potential%
\begin{equation}
V_{\func{Re}}\left( x\right) =-g^{2}\left( x\right) -\frac{1}{2}U\left(
x\right) U^{\prime \prime }\left( x\right) -\frac{1}{4}U^{\prime 2}\left(
x\right) +\varepsilon ,  \tag{26}
\end{equation}%
where $\varepsilon $ is some constant of integration. In consequence, using
Eqs.(22--24), we obtain the potential%
\begin{equation}
V\left( x\right) =-g^{2}\left( x\right) -iU\left( x\right) g^{\prime }\left(
x\right) -\frac{1}{2}U\left( x\right) U^{\prime \prime }\left( x\right) -%
\frac{1}{4}U^{\prime 2}\left( x\right) +\varepsilon .  \tag{27}
\end{equation}

\section{Equivalence of Complementarity--Coordinate transformation}

In this section, we bring to the notion of the complementarity a
mathematical meaning by examining the way in which pseudo--Hermiticity
should map into weak pseudo--Hermiticity through the generating functions $%
F\left( x\right) $ and $f\left( x\right) $. In fact, it is well known from
Eqs.(19) and (24) that both generating functions belong to the same ordinary
space representation $\left\{ X\right\} $, then there must be a
transformation connecting them. For this reason, we assume that the required
transformations are concerned with coordinate transformations (or point
canonical transformations.)

In mathematical terms, a coordinate transformation $x\equiv x\left( \xi
\right) $ changes $F\left( x\right) $ into $f\left( \xi \right) $ in the
following way%
\begin{equation}
F\left( x\right) =\frac{G\left( x\right) }{2}\left[ \frac{U\left( x\right) }{%
G\left( x\right) }\right] ^{\prime }\quad \overset{x\equiv x\left( \xi
\right) }{\longrightarrow }\quad f\left( \xi \right) =\frac{\overline{U}%
^{\prime }\left( \xi \right) }{2},  \tag{28}
\end{equation}%
where $U\left( x\right) \equiv U\left[ x\left( \xi \right) \right] =%
\overline{U}\left( \xi \right) $.

An interesting way to solve this problem, that can be described within
coordinate transformation, is to build a differential equation from Eq.(19)
and assume that it is maintained invariant if one applies a coordinate
transformation. In fact, Eq.(19) can be expressed as%
\begin{equation}
U\left( x\right) \frac{dZ\left( x\right) }{dx}=2F\left( x\right) Z\left(
x\right) ,  \tag{29}
\end{equation}%
where $Z\left( x\right) =\frac{U\left( x\right) }{G\left( x\right) }$. It is
then obvious that whenever Eq.(29) holds for the set of functions (i.e. $%
U\left( x\right) $, $F\left( x\right) $ and $Z\left( x\right) $), similar
differential equation will holds for the transformed functions too (i.e. $%
\overline{U}\left( \xi \right) $, $\overline{F}\left( \xi \right) $ and $%
\overline{Z}\left( \xi \right) $) such as%
\begin{equation}
\overline{U}\left( \xi \right) \frac{d\overline{Z}\left( \xi \right) }{d\xi }%
=2\overline{F}\left( \xi \right) \overline{Z}\left( \xi \right) ,  \tag{30}
\end{equation}%
where $F\left( x\right) \equiv F\left[ x\left( \xi \right) \right] =%
\overline{F}\left( \xi \right) $; idem. for $Z\left( x\right) $. Therefore,
from Eq.(30), the mass function $U\left( x\right) $ is changed in the
following way%
\begin{equation}
U\left( x\right) \rightarrow \overline{U}\left( \xi \right) =U\left[ x\left(
\xi \right) \right] \frac{d\xi \left( x\right) }{dx}.  \tag{31}
\end{equation}

Let us introduce two new functions $\mathcal{R}\left( \xi \right) $ and $%
\mathcal{S}\left( \xi \right) $ related, respectively, to $\overline{Z}%
\left( \xi \right) $ and $\overline{F}\left( \xi \right) $ by%
\begin{eqnarray}
Z\left( x\right)  &\rightarrow &\overline{Z}\left( \xi \right) =Z\left[
x\left( \xi \right) \right] \mathcal{R}\left( \xi \right) ,  \TCItag{32.a} \\
F\left( x\right)  &\rightarrow &\overline{F}\left( \xi \right) =F\left[
x\left( \xi \right) \right] \mathcal{S}\left( \xi \right) .  \TCItag{32.b}
\end{eqnarray}

Substituting Eqs.(32.a--b) into Eq.(30) taking into account (31), we get%
\begin{equation}
U\left( x\right) \frac{dZ\left( x\right) }{dx}=2\left[ \mathcal{S}\left(
x\right) F\left( x\right) -U\left( x\right) \frac{d}{dx}\ln \sqrt{\mathcal{R}%
\left( x\right) }\right] Z\left( x\right) ,  \tag{33}
\end{equation}%
and by identifying it to Eq.(29), one obtain%
\begin{equation}
\mathcal{S}\left( x\right) F\left( x\right) =F\left( x\right) +U\left(
x\right) \frac{d}{dx}\ln \sqrt{\mathcal{R}\left( x\right) },  \tag{34}
\end{equation}%
which can be interpreted as a similarity transformation relating $F\left(
x\right) $ to $f\left( x\right) $; i.e.%
\begin{equation}
F\left( x\right) \rightarrow f\left( x\right) \equiv \mathcal{S}\left(
x\right) F\left( x\right) =F\left( x\right) +U\left( x\right) \frac{d}{dx}%
\ln \sqrt{\mathcal{R}\left( x\right) }.  \tag{35}
\end{equation}

In this light, let us redefine the coordinate transformation on $F\left(
x\right) $ following%
\begin{equation}
F\left( x\right) \rightarrow \overline{F}\left( \xi \right) =F\left[ x\left(
\xi \right) \right] \mathcal{S}\left( \xi \right) =F\left[ x\left( \xi
\right) \right] \frac{d\xi \left( x\right) }{dx},  \tag{36}
\end{equation}%
and from Eqs.(35) and (19), we get the identity%
\begin{equation}
f\left( \xi \right) \equiv F\left[ x\left( \xi \right) \right] \mathcal{S}%
\left( \xi \right) =\frac{G\left[ x\left( \xi \right) \right] \mathcal{S}%
\left( \xi \right) }{2}\left[ \frac{U\left[ x\left( \xi \right) \right] }{G%
\left[ x\left( \xi \right) \right] }\right] ^{\prime }.  \tag{37}
\end{equation}

Now in order to recover our result, we assume that the condition $G\left[
x\left( \xi \right) \right] \mathcal{S}\left( \xi \right) =1$ holds, and by
defining the generating function $G\left( x\right) $ as%
\begin{equation}
G\left[ x\left( \xi \right) \right] \equiv \mathcal{S}^{-1}\left( \xi
\right) =\frac{dx\left( \xi \right) }{d\xi },  \tag{38}
\end{equation}%
therefore Eq.(37) can be amply simplified, taking into consideration
Eq.(30), to%
\begin{eqnarray}
f\left( \xi \right) &\equiv &\frac{1}{2}\left[ U\left[ x\left( \xi \right) %
\right] \frac{d\xi \left( x\right) }{dx}\right] ^{\prime }  \notag \\
&=&\frac{\overline{U}^{\prime }\left( \xi \right) }{2}.  \TCItag{39}
\end{eqnarray}

This completes the proof and leads to the identity (28).

\section{Conclusion}

In this paper, we have proposed to give a mathematical meaning for the
notion of complementarity between the twin concepts of pseudo--Hermiticity
and weak pseudo--Hermiticity within the framework of coordinate
transformations, and as a consequence this has opened the way towards
understanding the complementarity. Our primary concern in our work implies
that all generating functions, whose the associated potentials are related
to the pseudo--Hermiticity and weak pseudo--Hermiticity, can be connected
into some generalized coordinate transformations.

As a concluding remark, we would like to point out the equivalence between
the complementarity and coordinate transformations is concerned by a
particular choice which the generating function $G\left( x\right) $ (i.e. $%
F\left( x\right) $) can take.

\end{document}